\documentclass[aps,floatfix,showpacs,showkeys,amsmath]{revtex4}
\usepackage{color,graphicx}
\usepackage{dcolumn}
\usepackage{bm}
\usepackage{epsfig}
\usepackage{supertabular}
\usepackage[bookmarksopen]{hyperref}
\usepackage{url}
\usepackage{graphicx}
\usepackage{float}
\usepackage[caption = false]{subfig}
\def\be {\begin{equation}}
\def\ee {\end{equation}}
\def\nn {\nonumber}
\def\bea {\begin{eqnarray}}
\def\eea {\end{eqnarray}}
\def\rpp{\rho\pi\pi}

\begin{document} 
 \title{\Large {\bf Spectral properties of $\rho$ meson  in a magnetic field }}
\renewcommand{\thefootnote}{\alph{footnote}}
\bigskip
\bigskip
\author{Snigdha Ghosh$^{a,c}$}
\email{snigdha.physics@gmail.com}
\author{Arghya Mukherjee$^{b,c}$}
\email{arghya.mukherjee@saha.ac.in}
\author{Mahatsab Mandal$^b$}
\email{mahatsab@gmail.com}
\author{Sourav Sarkar$^{a,c}$}
\email{sourav@vecc.gov.in}
\author{ Pradip Roy$^{b,c}$}
\email{pradipk.roy@saha.ac.in}
\affiliation{$^a$Variable Energy Cyclotron Centre
1/AF Bidhannagar, Kolkata 700 064,
India}
\affiliation{$^b$Saha Institute of Nuclear Physics, 1/AF Bidhannagar
Kolkata - 700064, India}
\affiliation{$^c$Homi Bhabha National Institute, Training School Complex, Anushaktinagar, Mumbai - 400085, India}

\begin{abstract}
We calculate the rho meson mass in a weak magnetic field using effective $\rho\pi\pi$ interaction. It is seen that both $\rho^0$ and $\rho^\pm$ masses decrease with the  
magnetic field in vacuum. $\rho$\,meson dispersion relation has  been calculated and shown to be different for $\rho^0$ and $\rho^\pm$. We
also calculate the $\rpp$ decay width and spectral functions of $\rho^0$ and $\rho^\pm$. The width is seen to decrease  with $eB$ and the spectral functions become narrower.
\end{abstract}
\maketitle

\section{Introduction}
Quantum chromodynamics(QCD) in the presence of magnetic field has gained a lot of research interests in recent years. Apart from being exciting for its own intricacy and 
subtleties, the underlying physics of this strongly interacting  matter under extreme conditions is  enriched with many remarkable  effects \cite{lectnote871}
like chiral magnetic effect\cite{nuclphy797,nuclphy803,prd78},magnetic catalysis\cite{nuclphy462} as well as inverse magnetic catalysis effect\cite{jhep1202},
 superconductivity of vacuum\cite{prd82,arxiv1208,prl106} and many more. It is also a remarkable fact that non-central heavy ion collisions at RHIC
and LHC do have the potential to provide the platform for their experimental verifications. In a non-central heavy-ion collision at LHC, magnetic fields 
of the order $eB\sim$15$m_\pi^2$ ($B\sim 5\times10^{15}$Tesla) can be achieved \cite{IntJmodphy24} which is, in fact, higher than the typical QCD scale i.e $eB\sim m_\pi^2$.
Though in heavy-ion experiments, the fields are produced for a very short interval of time, they are good  enough to substantially affect the strongly interacting fire-ball.
Moreover,in case of weak magnetic field limit the situation becomes almost analogous to the  magnetic fields present  inside magnetars which can be as high as
$eB\sim$1 MeV$^2$ \cite{astroj392}. It is to be noted  that the word \lq weak\rq ~is used to emphasize the dominance of QCD scale 
over the  $eB$ scale. Thus  systematic understanding of strong interaction with weak  magnetic field background  can also have  significant applications in physics of neutron stars
\cite{prl95,prd76,prl100,prl105,prd82faya} as well as some other topics of cosmology and early universe. In this context we  briefly recall
the proceedings in one of the aforementioned effects namely magnetic field induced superconductivity of vacuum. 

Though the existence of vacuum superconductor was first proposed a few years ago in ref \cite{prd82} with point-like vector mesons,recent researches
considering internal(quark) structures of the mesons kept on throwing new insights into this emerging phenomena. In ref \cite{prd82} it was shown 
that non-minimal coupling of $\rho$ mesons to the electromagnetic field could result in magnetic-field-induced  superconductivity of
the cold vacuum along the magnetic field direction. But due to the Vafa-Witten theorem \cite{nuclphy234} and QCD inequalities, Hidaka and Yamamoto 
concluded in ref \cite{prd87} that 
QCD vacuum structure can not be changed only by a magnetic field i.e magnetic-field-induced charged vector meson condensation is impossible. Soon 
after their work it has been argued in ref \cite{prd86} that $\rho^{\pm}$ condensation in  magnetic field background is consistent with the Vafa-Witten
theorem because of the existence of Higgs like mechanism and was supported a year later in ref \cite{plb721} where it was pointed out that the stronger version of 
the theorem \cite{prd87} was plagued with the prejudiced choice of a generating functional on symmetric vacuum ignoring the other possibilities
of non-symmetric vacua. However, the authors of ref \cite{prd87} also performed lattice QCD 
calculation in support of their conclusions. Interesting comment about that can be found in ref \cite{prd89} where it was argued that although 
the results of ref \cite{prd87} based on quenched lattice QCD simulation show vanishing correlation in large volume limit still that can not be 
a reason to conclude 
against condensation because of the inherent inhomogeneous nature of the condensate. For example, fig.4 of ref \cite{prd89} clearly demonstrates
the fact that  the vanishing of mass at the transition point  depends on  the order of the phase transition. In a careful investigation in the framework of SU(2) NJL 
model Liu et al. \cite{prd91} had pointed out that  as the estimated critical field for charged $\rho$ meson condensation is not  strong enough, one needs to take into 
account the contributions of higher landau levels as well,~considering  which the masses of charged $\rho$ mesons with $S_z=1$ and $S_z=-1$ do vanish 
 at $eB_c\sim 0.2$ GeV$^2$. However, in a very recent work in Hidden Local Symmetry approach in constant magnetic field \cite{arxiv1511}, it has been 
found that $\mathcal{O}(eB)^2$ corrections which are arising from the considerations of $\mathcal{O}(p^6)$ terms of derivative/chiral expansion, can in fact, change 
the trend of the effective mass from decreasing to increasing one resulting in absence of any massless limit point. Thus it is obvious that unanimous 
agreement on the existence of vacuum superconductor demands more research work in this field. In a recent work, the  pion mass and dispersion relations have been calculated 
in  \cite{prd93} with non-zero $eB$ in vacuum using an effective lagrangian (with pseudoscalar as well as pseudovactor pion-nucleon interactions). There it was shown that,
for pseudoscalar coupling pion effective mass significantly decreases with weak external magnetic field. However, for pseudovector coupling , only a modest increase was 
reported. Using the same methodology, we, in this work, investigate the problem of $\rho$ meson condensation with a phenomenological lagrangian under
the influence of a constant weak magnetic field in vacuum. The modification of dispersion relations due to finite temperature effects will be reported 
in our future work\cite{tobe}.

The paper is organized as follows. In Sec.II we discuss the formalism for calculating $\rho$ meson self-energy in the presence of weak magnetic field.  
Following similar approaches as in ref \cite{prd93}, we first define the scalar field Feynman propagators in constant external Abelian gauge field \cite{prd71} by 
Schwinger's proper time formalism \cite{pr82} and then calculate the effective mass upto one-loop order in self-energy. The results of our calculation are presented in Sec.III
in which  the effective mass variations with weak external field are presented followed by the dispersion relations. Finally in Sec.IV
we conclude with a brief summary and discussions.

\section{Formalism}
The self-energy resummed  $\rho$ meson  propagator satisfies the Dyson-Schwinger equation,
\bea
iD_{\mu\nu}(k) &=& iD_{\mu\nu}^0(k)+iD_{\mu\lambda}^0(k)(-i\Pi_{\rho}^{\lambda\sigma}(k))iD_{\sigma\nu}(k)\nonumber\\
(D_{\mu\nu})^{-1} &=& (D_{\mu\nu}^0)^{-1}-\Pi_{\mu\nu}\label{invprop},
\eea
where the bare propagator for the massive vector field  is given by 
\be
iD_{\mu\nu}^0=\frac{-i}{k^2-m_\rho ^2+i\epsilon}\left(g^{\mu\nu}-\frac{k^\mu k^\nu}{m_\rho^2}\right).
\ee
The pole of the effective propagator leads to the following dispersion relation
\bea
\mbox{det}\big[-(k^2-m_{\rho}^2)g^{\mu\nu}+k^\mu k^\nu -\Pi^{\mu\nu}_\rho\big]&=&0\label{dispersion}.
\eea

The exact form of the propagator of a charged scalar particle with mass $m$ and charge $e$ in presence of a constant magnetic field can be  written as \cite{prd71,pr82}
\bea
D_B(x',x'')&=&\phi(x',x'')\int\frac{d^4p}{(2\pi)^4}e^{-ip.(x'-x'')}D_B(p),\\
&&\hspace{-5cm}\mbox{where}\nn\\
iD_B(p)&=&\int_{0}^{\infty}\frac{ds}{cos(eBs)}e^{is\left[p_{||}^2-p_{\perp}^2\big(\mbox{tan}(eBs)/eBs\big)-m^2+i\epsilon\right]}\label{momspace}\\
&&\hspace{-5cm}\mbox{and}\nn\\
\phi(x',x'')&=&\mbox{exp}\Big[ie\int_{x''}^{x'}dx_\mu A^\mu(x)\Big].
\eea
As the phase factor of the Schwinger's propagator is independent of the path, the overall phase of the one loop self-energy involving two scalar particles becomes unity. Thus
we can work in momentum space representation of the scalar propagator as given in eqn.(\ref{momspace}). In this paper, we use the following convention : $g^{\mu\nu}$ is 
decomposed into two parts as $g^{\mu\nu}=g^{\mu\nu}_{||}-g^{\mu\nu}_{\perp}$, where $g^{\mu\nu}_{||}=\mbox{diag}(1,0,0,-1)$ and $g^{\mu\nu}_{\perp}=\mbox{diag}(0,1,1,0)$. 
Similarly a general four vector can be written as $q^{\mu}=q_{||}^{\mu}+q_{\perp}^{\mu}$ with $q_{||}^2=q_{0}^2-q_{3}^2$ and $q_{\perp}^2=q_{1}^2+q_{2}^2$. Natural units 
will be used through out the paper. From now on, the $i\epsilon$ term in the propagator will not be explicitly written and will be taken care of at the end of the
calculation. 
The exact propagator in the external magnetic field can be written as a series in powers of $eB$ \cite{prd71}. As we are interested in weak field regime, keeping only the 
lowest order terms we get 
\bea
iD_B(p)\xrightarrow{eB\rightarrow 0}\frac{i}{p_{||}^2-p_{\perp}^2-m^2}\Big[1-\frac{(eB)^2}{(p_{||}^2-p_{\perp}^2-m^2)^2}
-\frac{2(eB)^2 p_{\perp}^2}{(p_{||}^2-p_{\perp}^2-m^2)^3}\Big].
\eea

\begin{figure}[h!]
\captionsetup[subfigure]{oneside,margin={0in,0in}}
\subfloat[]{\hspace{-0.3cm}\includegraphics[width = 3in]{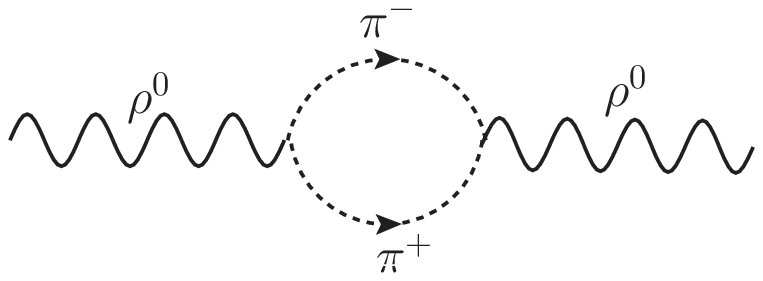}} \hspace{1cm}
\subfloat[]{\hspace{-0.3cm}\includegraphics[width = 3in]{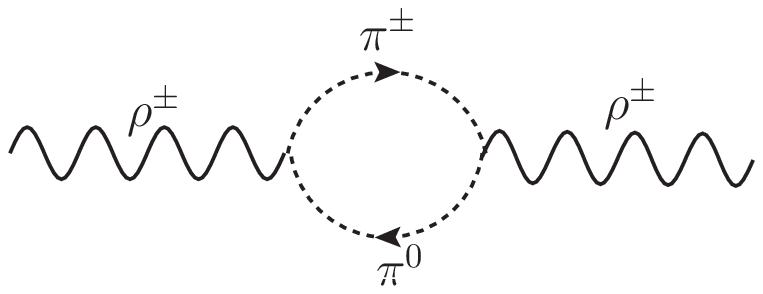}}\\  \subfloat[]{\includegraphics[width = 2in]{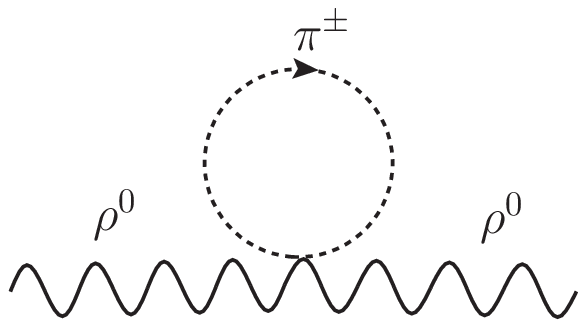}}
\hspace{3.5cm}
\subfloat[]{\includegraphics[width = 2in]{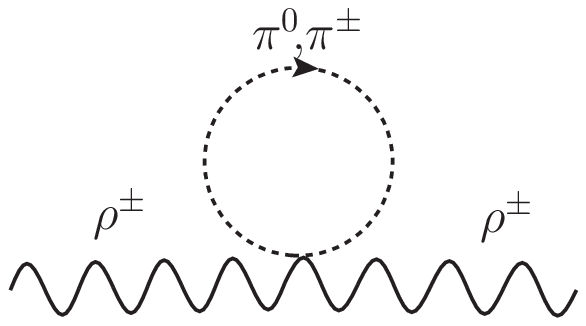}} 
\caption{Feynman Diagrams for the one-loop self energy of $\rho$}
\label{fig:1}
\end{figure}
The  phenomenological lagrangian  corresponding to $\rho\pi\pi$ interaction  can be written as  \cite{nuclphyA634} 
\be
\mathcal{L_{\rho\pi\pi}}=-g_{\rho\pi\pi}\mbox{\boldsymbol{$\rho$}}^{\mu}.(\mbox{\boldsymbol{$\pi$}}\times\partial_{\mu}\mbox{\boldsymbol{$\pi$}})+\frac{1}{2}g_{\rho\pi\pi}^{2}
(\mbox{\boldsymbol{$\rho^\mu$}}\times\mbox{\boldsymbol{$\pi$}}).(\mbox{\boldsymbol{$\rho_\mu$}}\times\mbox{\boldsymbol{$\pi$}}),
\ee
where the boldfaced $\rho$ and $\pi$ indicate that they are isovectors. Expanding the lagrangian with  complex pseudo-scalar and vector fields,
one can easily  get the  possible one-loop self-energy diagrams as shown in fig.1. In the following calculation we ignore the mass difference between neutral $\pi$ meson 
and the charged $\pi$ and denote the pion mass as $m_{\pi}$. Sometimes the superscript indices denoting neutral and charged $\rho$  will be written downstairs for 
aesthetic reasons. 

The part of the interaction  lagrangian which is responsible for the neutral $\rho$ meson self-energy can be explicitly written as  
 \be
 \mathcal{L}_{\rho_0} = i g_{\rho\pi\pi}[\rho_{0}^{\mu}(-\pi^{-}\partial_{\mu}\pi^{+}+\partial_{\mu}\pi^{-}\pi^{+})]+g_{\rho\pi\pi}^{2}\rho_{0}^{2}\pi^{-}\pi^{+}.
 \ee
 From which one can find the vertex factor for sub-diagram (a) to be $\Gamma_a^{\mu}=-ig_{\rho\pi\pi}(2p+k)^{\mu}$ and that for sub-diagram (c) to be 
 $\Gamma_c=i2!g^2_{\rho\pi\pi}$. 
 One loop Self-energy for $\rho_0$ meson is given by 
 \bea
 -i\Pi^{\mu\nu}_{\rho_0}&=&\int\frac{d^4p}{(2\pi)^4}\big[\Gamma^{\mu}_a \,iD_B(p+k)\,\Gamma^{\nu}_a\,iD_B(p)+g^{\mu\nu}\Gamma_c iD_B(p)\big]\nn\\
 \Pi^{\mu\nu}_{\rho_0}  &=& i g^2_{\rho\pi\pi}\int\frac{d^4p}{(2\pi)^4}\big[(2p+k)^{\mu}(2p+k)^{\nu}D_B(p+k)D_B(p)-2g^{\mu\nu}D_B(p)\big],
 \eea
 where 
\be
D_B(p) = D_0(p) -(eB)^2\frac{p^2_{||}+p^2_\perp-m^2_\pi}{(p^2-m^2_\pi)^4},
\ee
and $D_0(p) = [p^2-m^2_\pi]^{-1}$ is the free scaler propagator.
 
 It is to be noted here that, as the vacuum part and the magnetic correction  term  of the propagator are  additive , the one loop self-energy of $\rho_0$ can be decomposed into two parts  as 
 \bea
 \Pi^{\mu\nu}_{\rho^0} &=&  \Pi^{\mu\nu}_{\rho^0}(eB=0)+ \Pi^{\mu\nu}_{\rho^0}(eB\ne0).
\eea
This is true for the charged $\rho$ mesons as well.  After dimensional regularization and renormalization, the vacuum part of the self energy 
which is finite and scale dependent, can be written as  \cite{quig,nuclphyB357} 
\bea
 \Pi^{\mu\nu}_{\rho^0}(eB=0)&=&-(g^{\mu\nu}-\frac{k^\mu k^\nu}{k^2})\Pi_{\rm vac}(k^2)\nn\\
\mbox{with}\hspace{2cm} \Pi_{\rm vac}(k^2)&=&\frac{1}{3}\frac{g_{\rpp}^2}{16\pi^2}k^2\Big[\Big(1-\frac{4m_\pi^2}{k^2}\Big)^\frac{3}{2}\mbox{ln}\Big(
\frac{\sqrt{1-4m_\pi^2/k^2}+1}{\sqrt{1-4m_\pi^2/k^2}-1}\Big)
+\frac{8m_\pi^2}{k^2}+K\Big],
\eea
 where $K$ contains the mass scale and can be fixed by an additional condition based on physical grounds. Although $K$ is a constant in case of pure vacuum, however
in the presence of magnetic field, the additional condition makes it in principle a function of $eB$ as will be discussed in the next section. We denote  
the renormalized  mass of $\rho$ meson as $m_\rho$ whereas the magnetic field dependent effective mass is denoted as $m^{\ast}_\rho$.

In this work we are mainly concerned with the $eB$ dependent one loop self-energy up to $\mathcal{O}((eB)^2)$ which
is reasonable in the weak field regime. 
After  plugging in the propagators explicitly, the  expression for the magnetic field dependent part of  neutral $\rho$ meson self-energy becomes 
\bea
\Pi^{\mu\nu}_{\rho^0}(eB\ne0)&=&-i\,(eB)^2\,g^2_{\rho\pi\pi}\int\frac{d^4p}{(2\pi)^4}\Bigg[ (2p+k)^{\mu}(2p+k)^{\nu}\nn\\
&\times&\Bigg\{\frac{p^2_{||}+p^2_{\perp}-m^2_{\pi}}{[p^2-m^2_\pi]^4[(p+k)^2-m^2_\pi]}+\frac{(p+k)^2_{||}+(p+k)^2_{\perp}-m^2_{\pi}}{[p^2-m^2_\pi][(p+k)^2-m^2_\pi]^4}
\Bigg\}-2g^{\mu\nu}\frac{p^2_{||}+p^2_{\perp}-m^2_\pi}{(p^2-m^2_\pi)^4}\Bigg]\label{form}.\nn\\
\eea

Thus, we have altogether three integrals. Standard Feynman parametrization technique can be applied to these 
integrals one by one. Starting with the first one we find 
     
\bea
&&\int\frac{d^4p}{(2\pi)^4} (2p+k)^{\mu}(2p+k)^{\nu}\frac{p^2_{||}+p^2_{\perp}-m^2_{\pi}}{[p^2-m^2_\pi]^4[(p+k)^2-m^2_\pi]}\nn\\
&=&\int_0^1 dx\,4(1-x)^3 \int\frac{d^4p}{(2\pi)^4} \frac{1}{(p^2-\Delta)^5}\nn\\
&\times&\Bigg[ 4\big[p^\mu_{||}p^\nu_{||}+p^\mu_{\perp}p^\nu_{\perp}\big]\big[p^2_{||}+p^2_{\perp}+x^2(k^2_{||}+k^2_{\perp})\big]
+(2x-1)^2k^\mu k^\nu\big[p^2_{||}+p^2_{\perp}+x^2(k^2_{||}+k^2_{\perp})\big]\nn\\
&+&4x(2x-1)\big[k^\mu\{p^\nu_{||}\,(p_{||}\cdot k_{||})-p^\nu_{\perp}\,(p_{\perp}\cdot k_{\perp})\}+
k^\nu\{p^\mu_{||}\,(p_{||}\cdot k_{||})-p^\mu_{\perp}\,(p_{\perp}\cdot k_{\perp})\}\big]\nn\\
&-&m^2_{\pi}\big[4p^\mu p^\nu+(2x-1)^2k^\mu k^\nu\big]\Bigg].
\label{eqnappendics}
\eea
It is worth mentioning that out of all the terms that emerge after the change of variable $p\rightarrow p-xk$, the bracketed terms are the only  non-vanishing ones. Necessary
identities for all the momentum integrations are given in the appendix. Interestingly, one can skip  the tenure of 
the calculation for the second integral by noticing that $p\leftrightarrow p+k$ transforms it exactly to the 
first one. Combining  the contribution from the third integral with the first two, we find the following structure for the one-loop self energy of neutral $\rho$ meson :  
\bea
\Pi_{\rho^0}^{\mu\nu} &=& \Pi^{\mu\nu}_{\rho^0}(eB=0) + A_0k^\mu k^\nu + B_0g^{\mu\nu}+C_0g_\perp^{\mu\nu}+D_0\left(k^\mu k_\perp^\nu + k^\nu k_\perp^\mu\right)
\label{selfenergy}
\eea
with the structure functions given  as follows :
\bea
A_0 &=& \frac{2}{3}\frac{g_{\rpp}^2(eB)^2}{16\pi^2}\int_{0}^{1}dx(1-x)^3(1-2x)\left[\left(\frac{(1-2x)(m_{\pi}^2-2k_\perp^2x^2-x^2k^2)}{\Delta^3} \right) -\frac{2x}{\Delta^2}  \right] \\
B_0 &=& \frac{1}{3}\frac{g_{\rpp}^2(eB)^2}{16\pi^2}\left[\frac{1}{m_{\pi}^2}-\int_{0}^{1}dx(1-x)^32\left(\frac{1}{\Delta}+\frac{m_{\pi}^2-2k_\perp^2x^2-x^2k^2}{\Delta^2} \right)\right] \\
C_0 &=& - \frac{4}{3}\frac{g_{\rpp}^2(eB)^2}{16\pi^2}\int_{0}^{1}dx \left[\frac{(1-x)^3}{\Delta}\right] \\
D_0 &=& \frac{4}{3}\frac{g_{\rpp}^2(eB)^2}{16\pi^2}\int_{0}^{1}dx \left[\frac{x(1-x)^3(1-2x)}{\Delta}\right]
\eea
where  $\Delta=x(x-1)k^2+m^2_{\pi}-i\epsilon$. Note the $i\epsilon$ term in the expression which takes into account its presence in all the Feynman propagators, not 
explicitly mentioned earlier. In certain kinematic domain (like $k^2\ge4m_{\pi}^2$ where the Unitary cut begins), the structure constants  can have significant real
and imaginary parts.

In case of  $\rho^\pm$ mesons the contributing interaction lagrangian is given by 
 $$\mathcal{L_{\rho^\pm}}=\rho^\pm_\mu(\pm\pi^\mp\partial^\mu\pi_0\mp\pi^0\partial^\mu\pi^\mp)+g_{\rho\pi\pi}^{2}\rho^{-}(\pi_{0}^{2}+\pi^{-}\pi^{+})\rho^{+}.$$
Following  similar procedure, magnetic field dependent self-energy of  $\rho^\pm$ up to   $\mathcal{O}(eB)^2$ can be written as 
 \bea
 \Pi^{\mu\nu}_{\rho^\pm}&=& i g^2_{\rho\pi\pi}\int\frac{d^4p}{(2\pi)^4}\big[(2p+k)^{\mu}(2p+k)^{\nu}D_B(p)D_0(p+k)-g^{\mu\nu}(2D_0(p)+2D_B(p))\big]\nn\\
 &=&-i\,(eB)^2\,g^2_{\rho\pi\pi}\int\frac{d^4p}{(2\pi)^4}\Bigg[ (2p+k)^{\mu}(2p+k)^{\nu}
\frac{p^2_{||}+p^2_{\perp}-m^2_{\pi}}{[p^2-m^2_\pi]^4[(p+k)^2-m^2_\pi]}-g^{\mu\nu}\frac{p^2_{||}+p^2_{\perp}-m^2_\pi}{(p^2-m^2_\pi)^4}\Bigg]
\eea
 From this structure one can straightforwardly conclude that 
\bea
\Pi_{\rho^{\pm}}^{\mu\nu} &=& \Pi^{\mu\nu}_{\rho^{\pm}}(eB=0) + A_{\pm}k^\mu k^\nu + B_{\pm}g^{\mu\nu}+C_{\pm}g_\perp^{\mu\nu}+D_{\pm}\left(k^\mu k_\perp^\nu + k^\nu k_\perp^\mu\right),
\eea
 where  structure functions  $A_{\pm},B_{\pm},C_{\pm},D_{}$ are nothing but half of the $A_0,B_0,C_0$ and $D_0$ respectively. 
 
The  decay width of  $\rho\rightarrow\pi\pi$ in  presence of magnetic field  is related to the imaginary part of the self-energy as \cite{nuclphyB357} 
\bea
\Gamma_{\rho}(eB)&=&\frac{\mbox{Im}\Pi(k_0= m^\ast ,eB)}{m^\ast}\label{decay}
\eea

where $\Gamma_{\rho}$ is defined in the rest frame of $\rho$ with $\Pi=\frac{1}{3}\Pi_{\mu}^\mu$  and $m^\ast$ is the solution of the equation
\bea
m^\ast{}^2-m_{\rho}^2+\mbox{Re}\Pi(k_0= m^\ast ,eB)&=& 0\label{mast}
\eea

For a given value of $eB$, $m^\ast$ gives the maximum of the spectral function which is  defined as 
\bea
\rho(k_0,eB)&=&\frac{\mbox{Im}\Pi}{(k_0^2-m_{\rho}^2+\mbox{Re}\Pi)^2+(\mbox{Im}\Pi)^2}
\eea
Note that, being a function of both $\mbox{Re}\Pi$ and $\mbox{Im}\Pi$ it carries all the essential features of the self-energy.

\section{Results}
\begin{figure}[H]
\subfloat[]{\includegraphics[width = 2.5in,angle=-90]{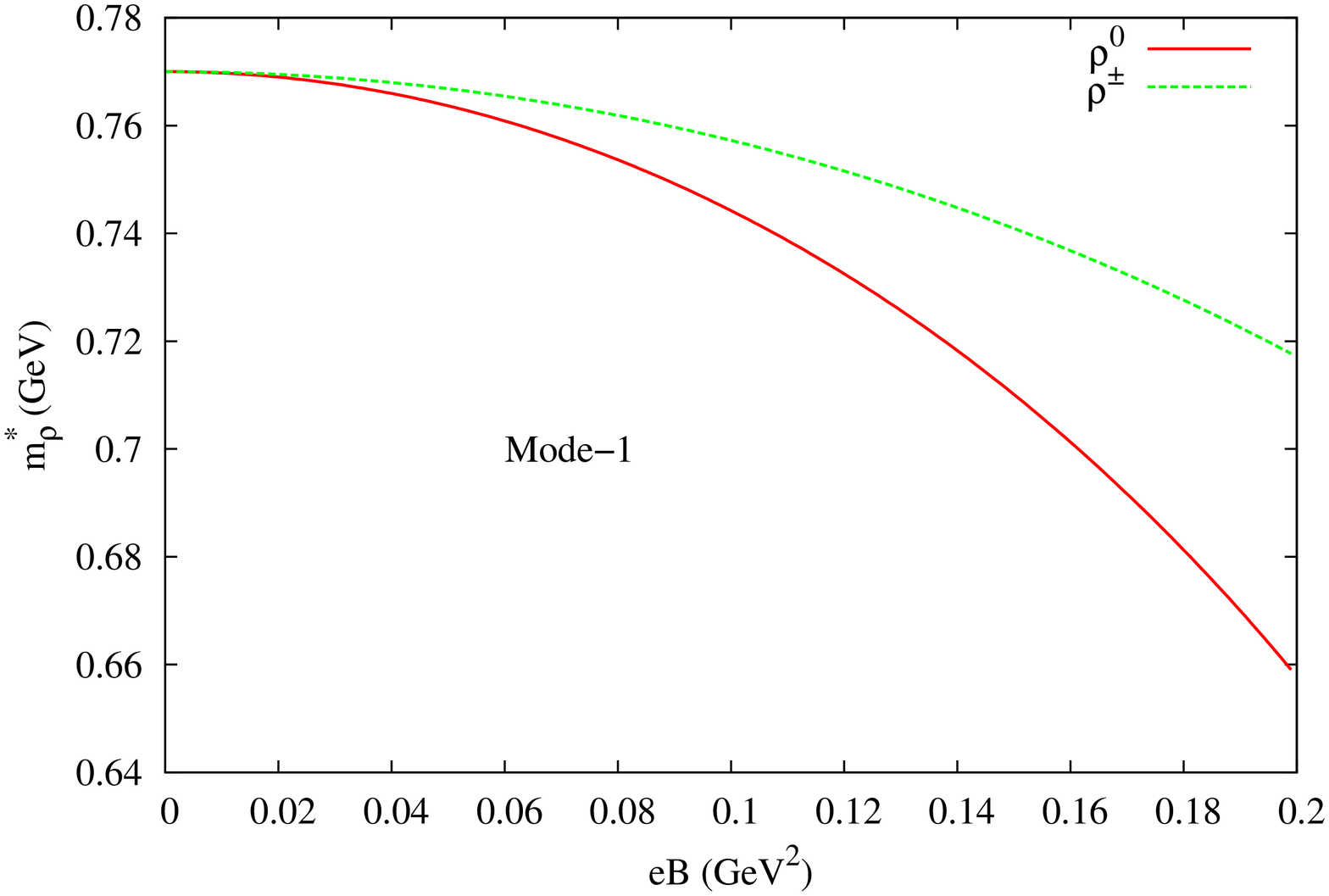}} 
\subfloat[]{\includegraphics[width = 2.5in,angle=-90]{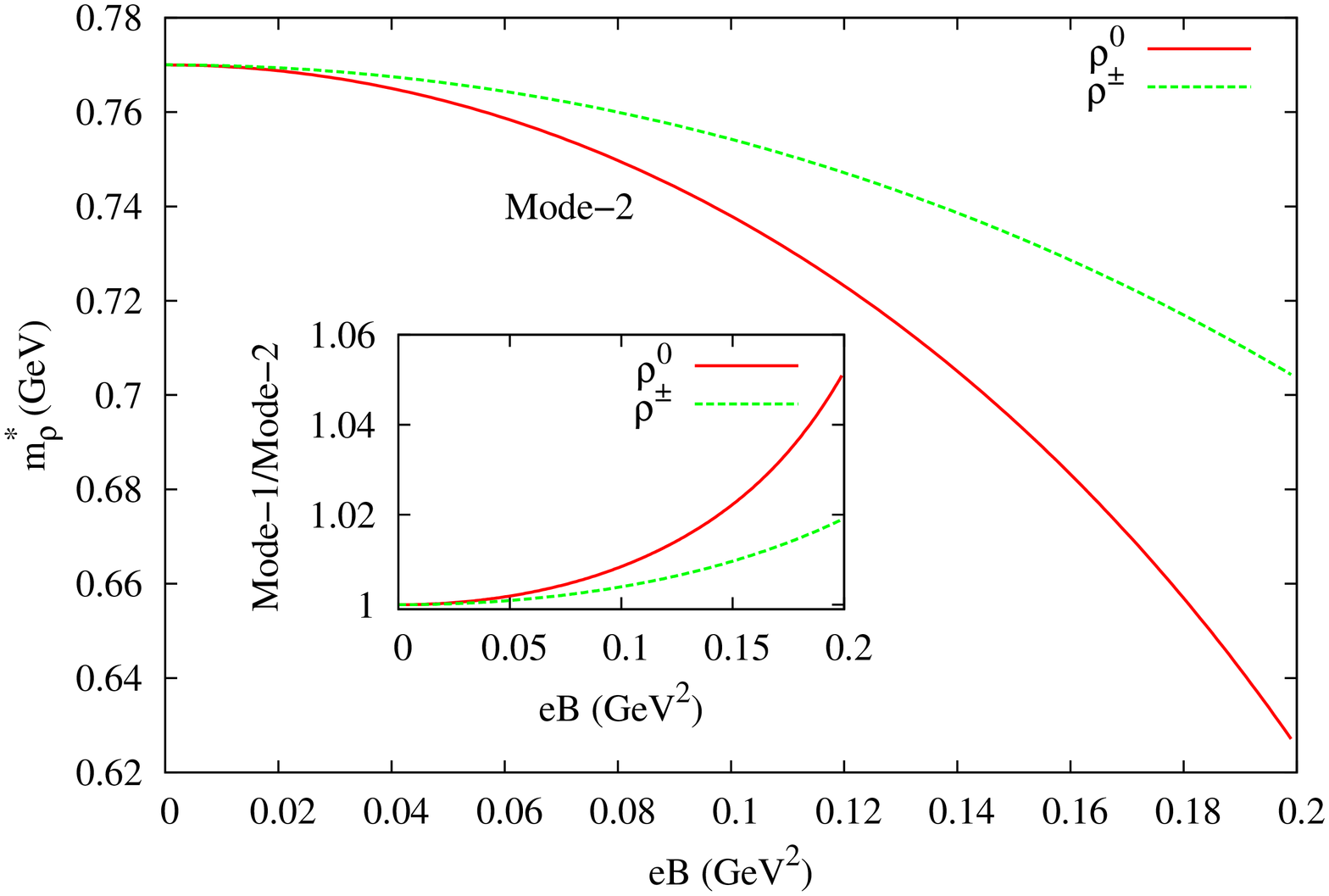}}
\caption{Effective mass variations for $\rho\pi\pi$ coupling as a function of $eB$. Both of the modes show decrease in  effective mass  of $\rho^0$ 
and $\rho^{\pm}$ with the increasing external field. The ratio plotted in the inset demonstrates the difference between the two modes.  }
\label{fig2}
\end{figure}

In this section we present the numerical results for the variation of the effective mass with weak external magnetic field. We consider the strength of the external 
field to be much less than the square of the $\rho$ meson mass i.e $eB\ll m^2_\rho$. In our numerical calculation, we have taken the coupling constant $g_{\rho\pi\pi}$ 
as 6.03 which can be obtained by using  the decay width of $\rho\rightarrow\pi\pi$ as 150 MeV. The mass of the pion  is taken as 0.14 GeV.To get the  
effective mass numerically, we  set the external three momentum of
Eq.(\ref{dispersion}) to zero and obtain four mass relations given by 
\bea
-m_{\rho}^2+A k_0^2+B&=&0\\
k_0^2-m_{\rho}^2-\Pi_{\rm vac}(k_0^2)+ B-C&=&0\\
k_0^2-m_{\rho}^2-\Pi_{\rm vac}(k_0^2)+ B-C&=&0\\
k_0^2-m_{\rho}^2-\Pi_{\rm vac}(k_0^2)+ B&=&0
\eea
 It must be noted here that for a given value of the parameter $eB$ 
each of the equations possess two unknowns, $k_0$ and the scale hidden in $\Pi_{\rm vac}$. It might seem that the scale 
is already fixed by the condition employed at $eB=0$ which is ${\rm Re}\Pi_{\rm vac}(k^2=m_\rho^2)=0$. But the physical mass in presence of magnetic field is $m^\ast_\rho$ 
and not the vacuum mass  $m_\rho$. Thus we must choose a more general condition ${\rm Re}\Pi_{\rm vac}(k^2=m_\rho^{\ast2})=0$ which correctly reproduces
the vacuum results in absence of $eB$. Using the above condition we get the following mass relations 

\bea
-m_{\rho}^2+Am_{\rho}^{\ast 2}+B&=&0\\
m_{\rho}^{\ast 2}-m_{\rho}^2+ B-C&=&0\\
m_{\rho}^{\ast 2}-m_{\rho}^2+ B-C&=&0\\
m_{\rho}^{\ast 2}-m_{\rho}^2+ B&=&0
\eea
where the subscripts of $A,B,C$ and $m^{\ast}_\rho$ are chosen accordingly. 
Out of the four relations, the first one gives unphysical mode whereas the second and third one being same, are denoted as Mode-1 with the last one denoted as Mode-2. 
In both of the cases,
we find that the  effective mass of $\rho^0$ as well as $\rho^{\pm}$ decreases with $eB$  as shown in fig.\ref{fig2}.  
To understand the connection between the modes and the spin states 
explicitly, one should note that,  in the rest frame of a massive vector particle, the completeness relation  satisfied by the polarization vectors  is given as 
\bea
\sum^3_{s=1}\epsilon_s^\mu\epsilon_s^{\ast\nu}&=&-g^{\mu\nu}+u^\mu u^\nu
\eea
where 
\bea 
\epsilon_1^\mu({\bf k})&=&\frac{1}{\sqrt{2}}(0,1,i,0)\nn\\
\epsilon_2^\mu({\bf k})&=&\frac{1}{\sqrt{2}}(0,1,-i,0)\nn\\
\epsilon_3^\mu({\bf k})&=&(0,0,0,1)\nn\\
u^\mu&=&(1,0,0,0).
\eea
Using the completeness relation, the self-energy function in Eq.(\ref{selfenergy}) can be decomposed in terms of the projection operators  
$P_s^{\mu\nu}=-\epsilon_s^\mu\epsilon_s^{\ast\nu}$ and $u^\mu u^\nu$. Inverting Eq.(\ref{invprop}) one gets the one loop corrected propagator as
\bea 
D^{\mu\nu}&=&\frac{P_1^{\mu\nu}}{k_0^2-m_{\rho}^2-\Pi_{\rm vac}(k_0^2)+ B-C}+\frac{P_2^{\mu\nu}}{k_0^2-m_{\rho}^2-\Pi_{\rm vac}(k_0^2)+ B-C}\nn\\
&+&\frac{P_3^{\mu\nu}}{k_0^2-m_{\rho}^2-\Pi_{\rm vac}(k_0^2)+ B}+\frac{u^\mu u^\nu}{-m_{\rho}^2+A k_0^2+B}.
\eea
This form of the propagator simply indicates that  Mode-1 physically represents  the spin state $S_z=\pm1$ whereas 
$S_z=0$ is represented by Mode-2.
 
Although we started with the same physical mass for both $\rho^0$ and $\rho^\pm$ which is 770 MeV, in both  the modes, their effective masses vary differently 
showing faster decrease for $m^{\ast}_{\rho^0}$ compared to  $m^{\ast}_{\rho^{\pm}}$. However, if we compare the variations in the two modes by plotting the ratio of 
the effective masses as a function of $eB$ 
(shown in the inset), we observe a difference between them which is in fact, relatively more prominent in case of $\rho^0$.
  The decreasing nature indicates 
the possibility  of $\rho$ condensation for higher magnetic fields. It also indicates that the critical field for $\rho^0$  meson should  be smaller in magnitude compared
to that for charged $\rho$.  However, as we are working in weak field regime, the prediction about the critical field $eB_c$ is beyond 
the scope of our approximation. It is to be noted here that we find non-zero effective mass for $\rho^{\pm}$ even at $eB=0.2$ $\mbox{GeV}^2$ which differs from that 
 predicted in ref.\cite{prd91}. Comparing with lattice results in  ref.\cite{lattice}, we find that our results agree in case of $\rho^0$ with $S_z=0$ , $\rho^+$ with
 $S_z=+1$ and  $\rho^-$ with  $S_z=-1$. These are the states for which $\rho$ mass decreases with the magnetic field. In rest of the cases, 
 increase in $eB$ also increases the mass.
 
 \begin{figure}[h]
\captionsetup[subfigure]{oneside,margin={0in,0in}}
\subfloat[]{\includegraphics[width = 2.5in,angle=-90]{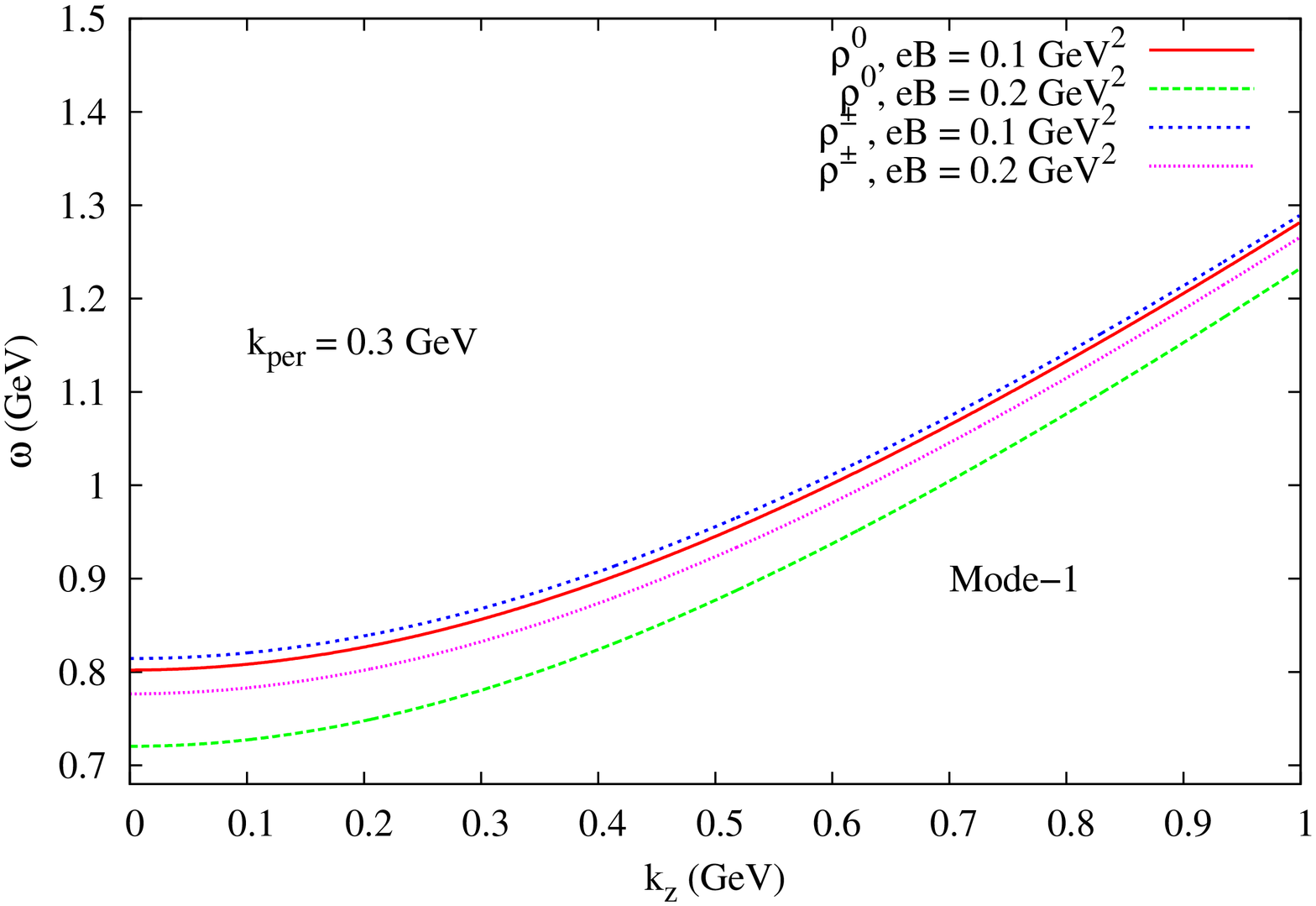}} 
\subfloat[]{\includegraphics[width = 2.5in,angle=-90]{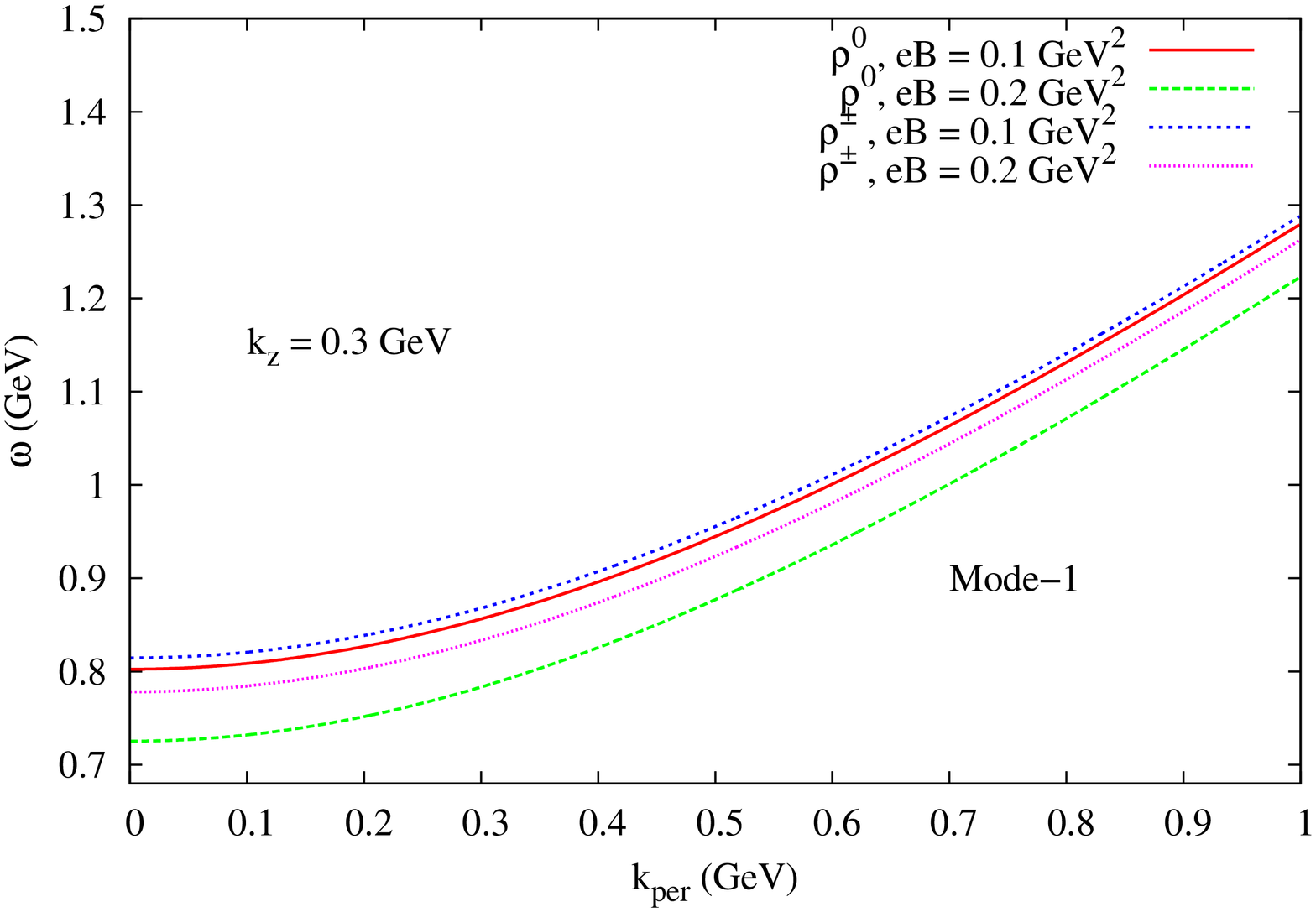}}\\
\subfloat[]{\includegraphics[width = 2.5in,angle=-90]{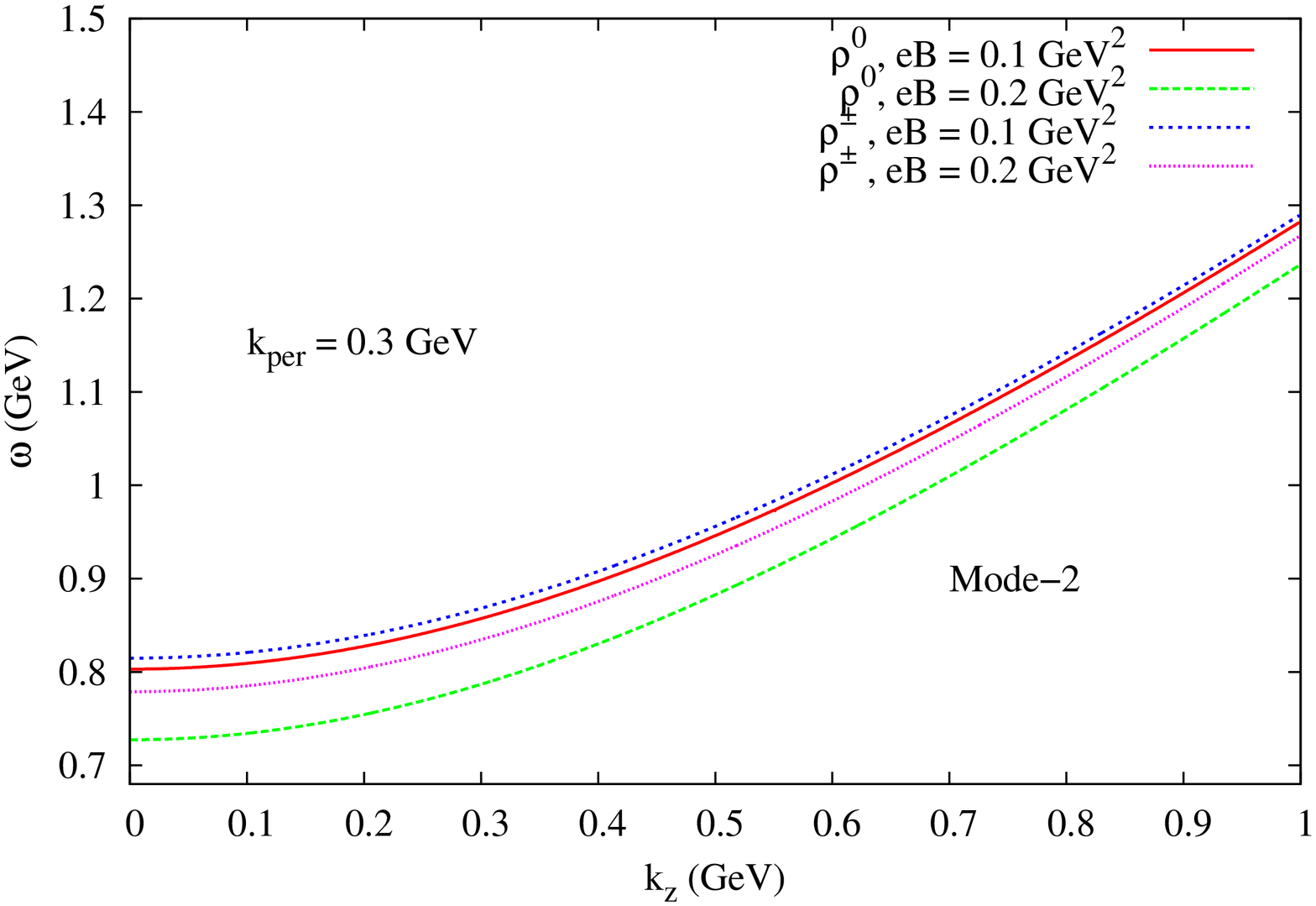}}
\subfloat[]{\includegraphics[width = 2.5in,angle=-90]{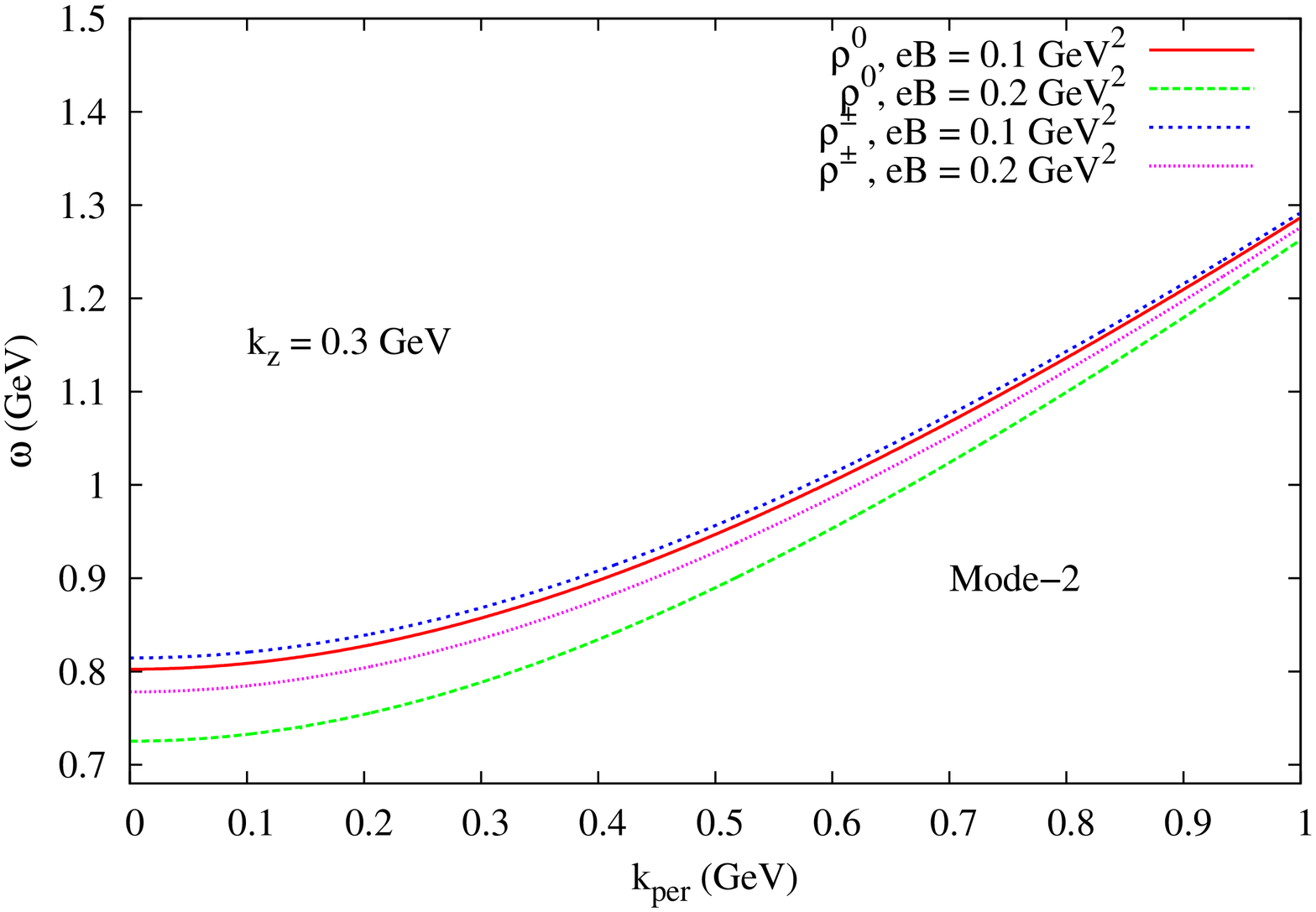}}\\
\subfloat[]{\includegraphics[width = 2.5in,angle=-90]{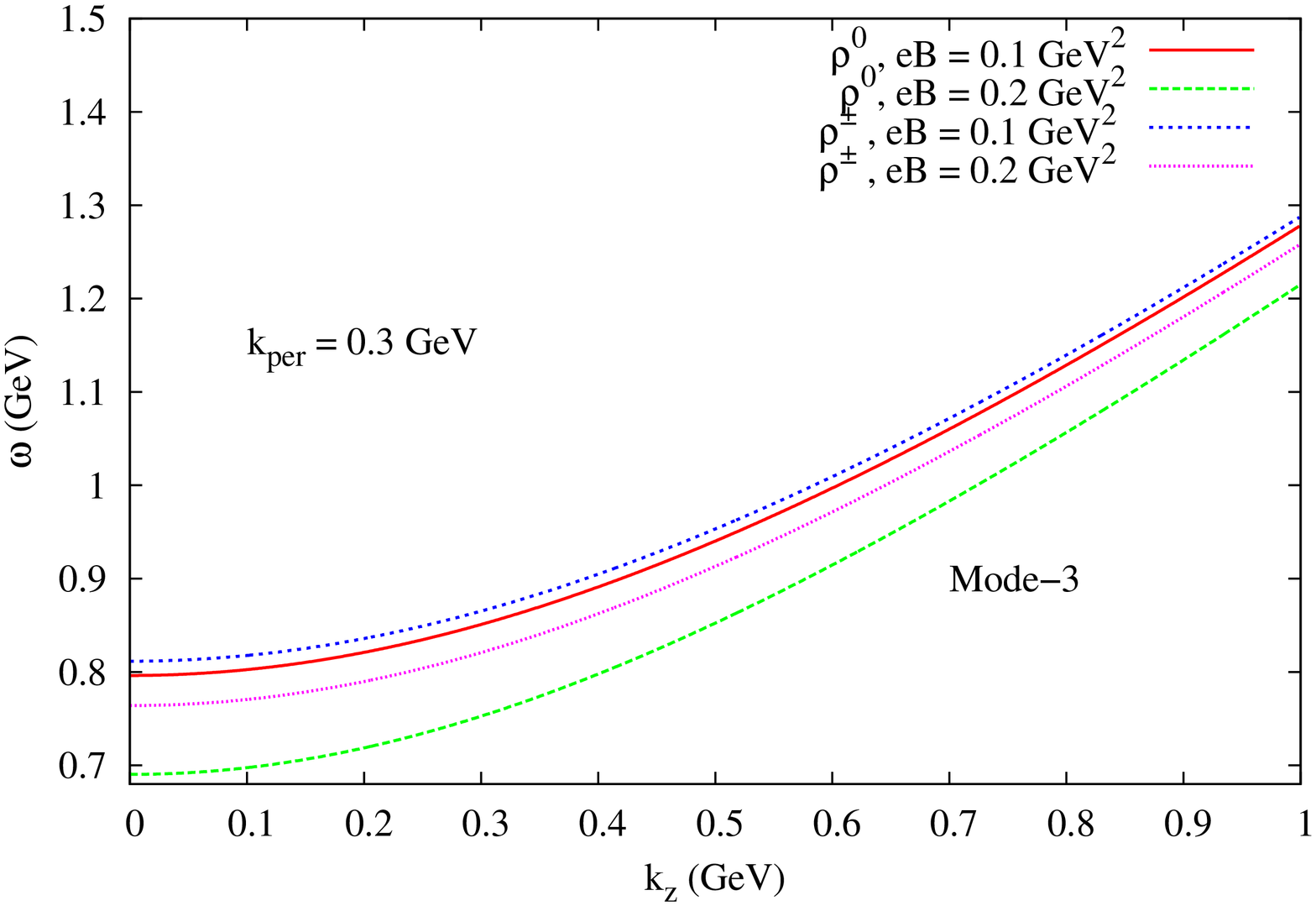}}
\subfloat[]{\includegraphics[width = 2.5in,angle=-90]{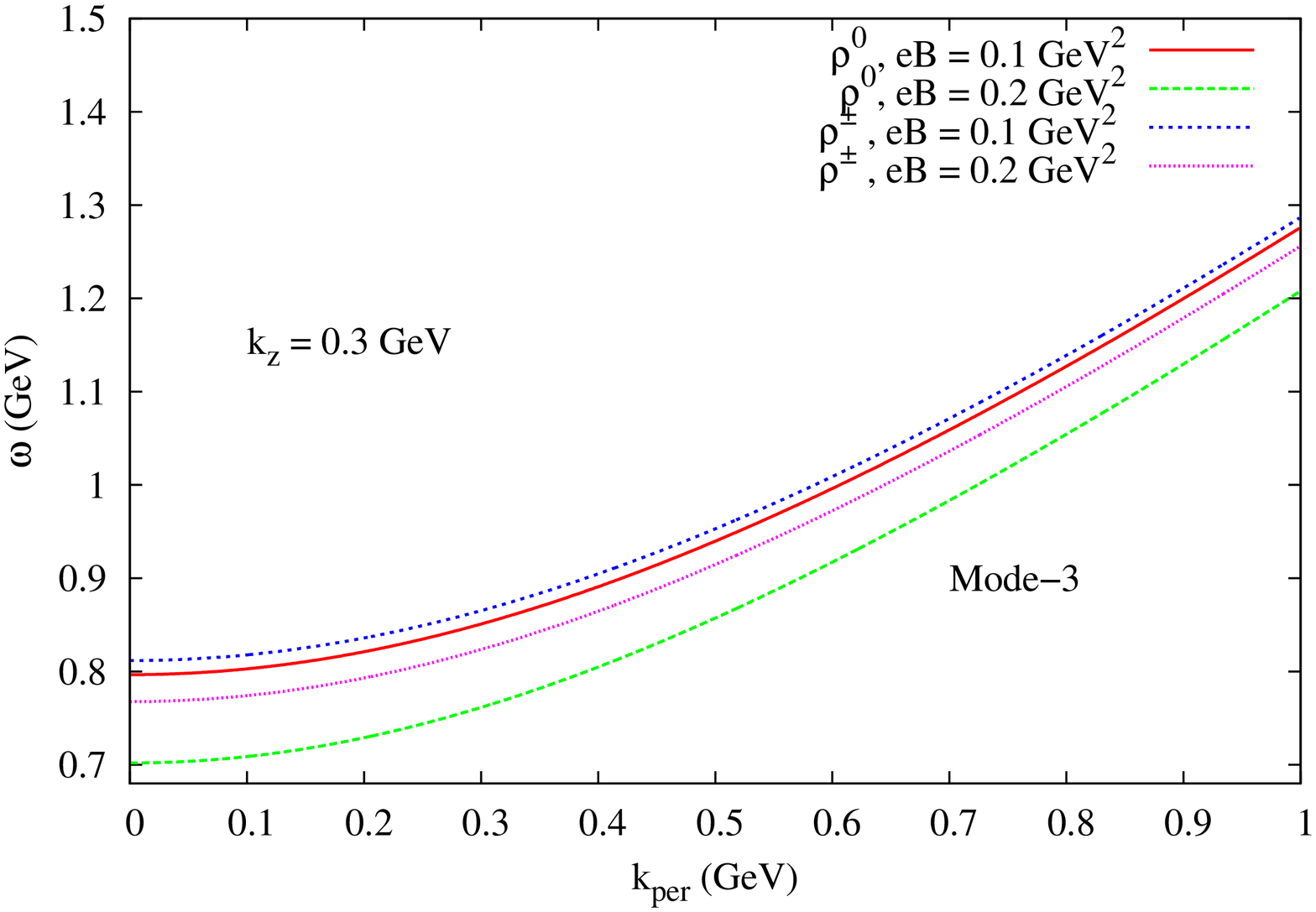}} \\
\caption{Dispersion relations of $\rho^0$ and $\rho^\pm$  for two different values of $eB$ in the weak field regime. Left panel shows the variation with $k_z$ for a 
fixed value of $k_{per}$. Right panel instead shows the dispersion as a function of $k_{per}$ keeping the $k_z$ fixed. }
\label{fig.4}
\end{figure}

In case of  dispersion relations, without any loss of generality we can reorient our axes such that $k_\perp^\mu=(0,k^1,k^2,0)$ 
becomes $k_\perp^\mu=(0,0,k_{per},0)$. Now, fixing the value of one of the independent variables in $k^\mu=(\omega,0,k_{per},k_z)$,
we can find the variation of $\omega$ with respect to the other. 
In fig.\ref{fig.4} the
 first column shows the variation of $\omega$ as a function of $k_z$ with $k_{per}=0.3$ GeV. The mode energy increases with the increase of longitudinal momentum in all the 
 three modes tending to coincide with the vacuum for higher $k_z$ values. This behaviour is plausible because, for  $k_z\gg eB$ the magnetic corrections do not contribute   
 significantly resulting in  light like dispersion. Similar behaviour can be observed from the second column where $k_{per}$ is varied keeping the longitudinal 
 momentum fixed at 0.3 GeV. With the increase of $eB$ from 0.1 to 0.2 $\mbox{GeV}^2$, one can notice the downward shift of the dispersion curves in all the three modes.

\begin{figure}[h!]
\captionsetup[subfigure]{oneside,margin={0in,0in}}
\subfloat[]{\includegraphics[width = 2.5in,angle=-90]{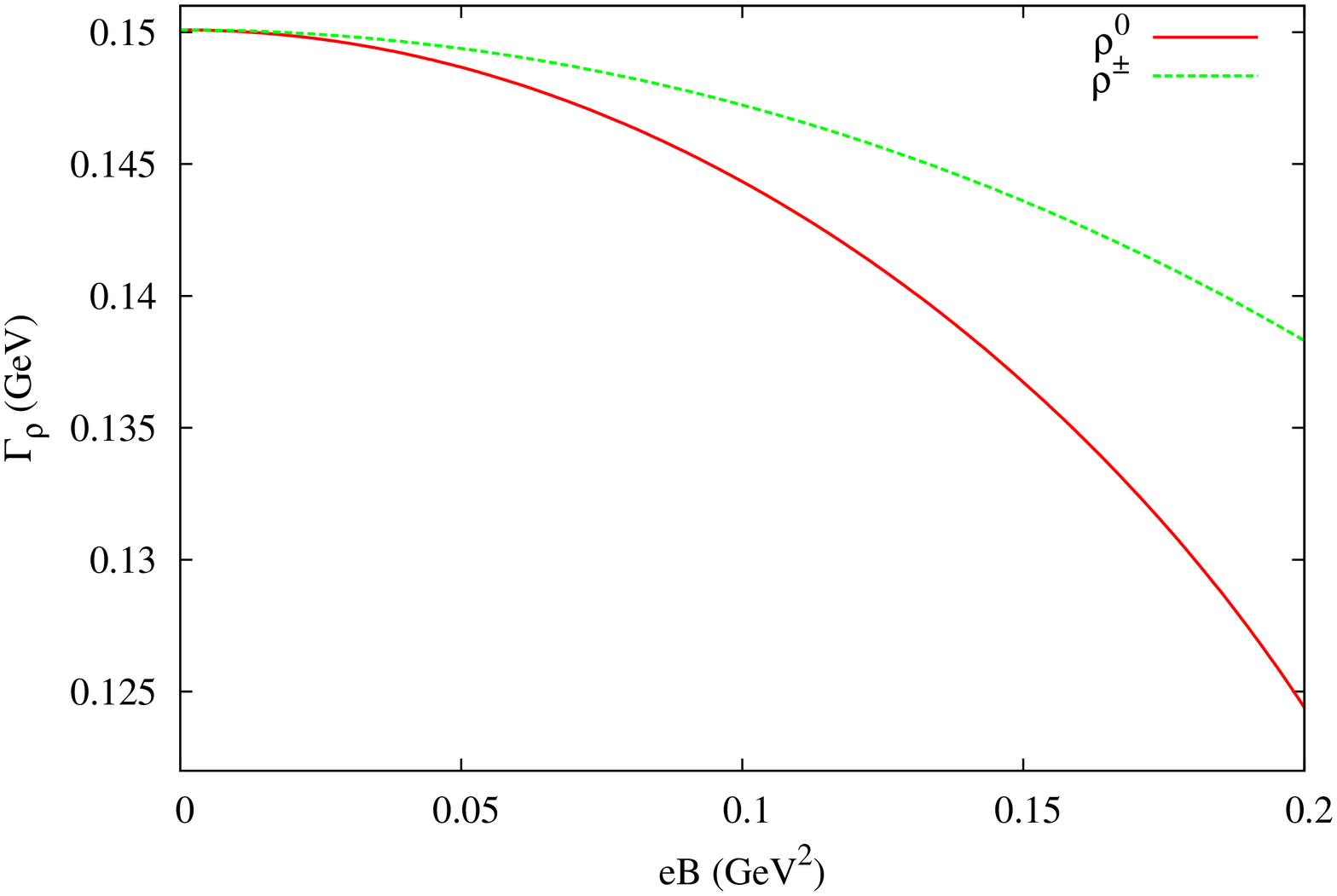}} 
\subfloat[]{\includegraphics[width = 2.5in,angle=-90]{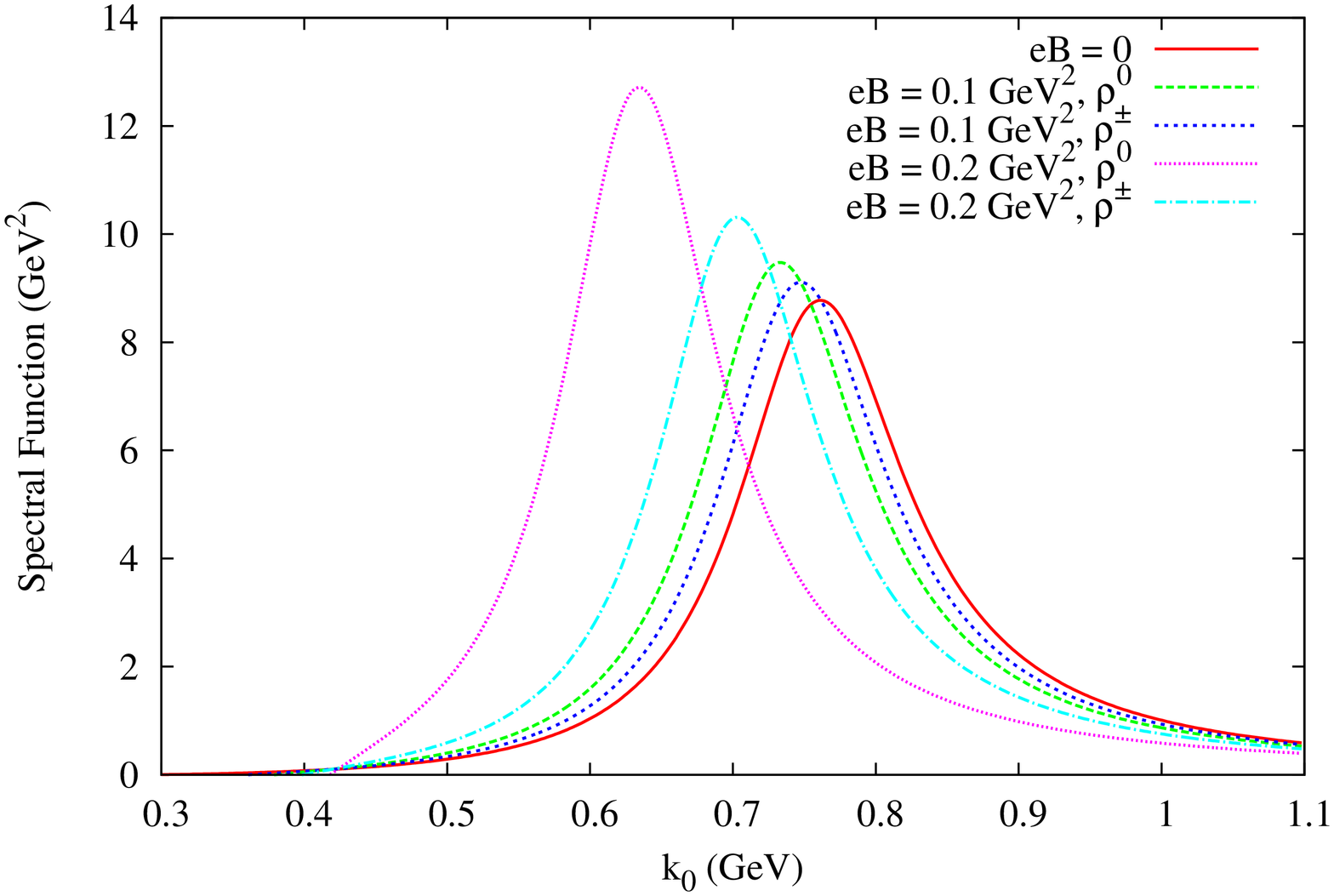}}\\
\caption{ Left panel shows the variation of  decay width  of $\rho$  mesons with external magnetic field. $\Gamma_\rho$ for both $\rho^0$ and $\rho^\pm$
decreases with magnetic field but remain finite even at maximum value of $eB=0.2~\mbox{GeV}^2$. Spectral functions are 
plotted in  the right panel for $eB=\{0,0.1,0.2\}~\mbox{GeV}^2$. It becomes narrower and taller as it shifts towards the lower values of $k_0$.}
\label{fig.5}
\end{figure}

 To calculate the decay width and  spectral function in the rest frame of $\rho$, we need to know the imaginary parts of $A,B$ and $C$.  The imaginary parts of those  
 structure functions can be  obtained analytically and  are given as follows
 \bea
 \frac{1}{2}\mbox{Im}[A] &=&-\frac{\pi}{2}\left[\frac{96 m_{\pi }^4}{\left(k_0^2\right){}^{5/2} \left(k_0^2-4 m_{\pi }^2\right){}^{3/2}}
 -\frac{32 m_{\pi }^2}{\left(k_0^2\right){}^{3/2} \left(k_0^2-4 m_{\pi }^2\right) {}^{3/2}}\right.\nn\\
 &&\left.+\frac{2}{\sqrt{k_0^2} \left(k_0^2-4 m_{\pi }^2\right){}^{3/2}}+\frac{24 m_{\pi }^2}{\left(k_0^2\right){}^{5/2} \sqrt{k_0^2-4 m_{\pi }^2}}-
 \frac{4}{\left(k_0^2\right)
 {}^{3/2} \sqrt{k_0^2-4 m_{\pi }^2}}\right](eB)^2\\
 -\frac{1}{2}\mbox{Im}[B]&=&-\frac{\pi}{2}\left[\frac{24 m_{\pi }^4}{\left(k_0^2\right){}^{3/2} \left(k_0^2-4 m_{\pi }^2\right){}^{3/2}}
 -\frac{6 m_{\pi }^2}{\sqrt{k_0^2} \left(k_0^2-4 m_{\pi }^2\right){}^{3/2}}
 +\frac{ 12\frac{m_{\pi }^2}{k_0^2}-4}{2\sqrt{k_0^2} \sqrt{k_0^2-4 m_{\pi }^2}}\right](eB)^2\\
 -\frac{1}{4}\mbox{Im}[C]&=&-\frac{\pi }{2}\left[\frac{ 12\frac{m_{\pi }^2}{k_0^2}-4}{2\sqrt{k_0^2} \sqrt{k_0^2-4 m_{\pi }^2}}\right](eB)^2
 \eea
 where the expressions are  scaled by the overall common factor $g_{\rpp}^2/48\pi^2$ for $\rho^0$ and half of that for $\rho^{\pm}$. Using the definition given in 
 Eq.(\ref{decay}) we obtain the decay width for $\rho\rightarrow\pi\pi$  as shown in the left panel of fig.\ref{fig.5}. It has been found that $\Gamma_\rho$ decreases 
 with  the external magnetic field both for $\rho^0$ and $\rho^\pm$. However, the rate of decrease being small, it
 
\noindent never vanishes even for $eB=0.2~\mbox{GeV}^2$ which we have taken to be the maximum limit of the external field as mentioned earlier. The fact indicates that, 
in the weak field limit, there exists suppression in the decay channel of 
$\rho\rightarrow\pi\pi$ but the prediction for complete blockage of the channel is beyond the scope of its applicability. 

Spectral functions are plotted in the right panel with two non-zero values of $eB$. As soon as the magnetic field is turned on, the vacuum spectral function 
splits into two, corresponding  to different self-energies of $\rho^0$ and $\rho^\pm$.  It is the interplay between $k_0$ and $eB$ dependencies of  Im$\Pi$ which makes the
spectral function narrower and taller as it shifts towards the condensation. The shift  is the manifestation of decreasing 
$m^\ast$ which is the $k_0$ value corresponding to  the maximum of the spectral function. These features of the spectral function are consistent with the qualitative 
 discussions given in \cite{prd82}. However, unlike \cite{prd82},  the shift for  $\rho^0$  in our case is more in comparison with that of $\rho^\pm$ which is expected
 from our  results of $m_\rho{}^\ast$ and $\Gamma_\rho$ .  
 
 \section{Summary and Conclusions}
 In this work we have investigated the charged and neutral $\rho$ meson condensation in external magnetic field. We restricted ourselves to the  weak field regime and 
  used the series form of the scalar propagator in magnetic background  taking into consideration the leading order magnetic correction term. Starting from the phenomenological
  lagrangian, 
 we explicitly calculated the magnetic correction to the one-loop self energy up to $\mathcal{O}(eB)^2$. Using the standard 
  Dyson-Schwinger equation, we have calculated  the effective mass variations for $\rho^0$ and $\rho^{\pm}$. In this case we find two independent physical modes both of 
  them showing decreasing nature   which 
  indicates the possibility of $\rho$ condensation. However, because of our restriction to the weak field regime, we can not predict the  exact value of critical
  field at which the  effective mass vanishes but we do find non-zero 
  $m^{\ast}_{\rho^{\pm}}$ for $eB=0.2$ GeV. The trend shows that critical 
  field for $\rho$ meson (both $\rho^0$ and $\rho^\pm$) will be higher than the prediction made in \cite{prd91}. Moreover, in our case $\rho^0$ mass falls faster 
  than $\rho^\pm$ 
  with increasing $B$ which differs from  the result of Liu et al \cite{prd91}. In addition to that we have also presented   the modified 
 dispersion relations of $\rho$ meson for  three distinct modes . Imaginary parts of the structure functions have been obtained analytically.  
  We have also explicitly calculated the spectral function  for the first time. The shift in the spectral function with the increasing magnetic field 
  is in agreement with what 
 has been anticipated in \cite{prd82} based on qualitative arguments. Another important difference of our work in comparison with that of ref.\cite{prd91} is that, in NJL model, the quarks are affected by the magnetic field. 
  That means the magnetically corrected propagators appearing in the self-energy should be fermionic propagators. But in our case, pionic fields contribute to the 
  $\rho$ self-energy.  Thus bosonic propagators contain the magnetic corrections. Now, the essential difference between the two is that in weak field  expansion,
  fermionic   propagators possess corrections of $eB$ order \cite{prd62} but in case of bosons, the leading order correction is $\mathcal{O}((eB)^2)$ \cite{prd71}.
  Thus, one can expect that no $eB$ order correction can be introduced with it.
  It is true that depending upon the interactions, even if one uses $\mathcal{O}(eB)$ corrected fermionic propagators,  leading order contribution to self-energy
  may not be of  $\mathcal{O}(eB)$\cite{prd93}. However, the discussions in 
  \cite{prd91} clearly indicates that at least in case of $m_{\rho^{\pm}}$ with $S_z=\pm1$, magnetic correction of $\mathcal{O}(eB)$ exists and it is 
  the only mode for which decrease in mass has been observed. Our spin decomposition in the rest frame  is similar to that of \cite{prd91}. We are also not considering 
  the LLL approximation. Thus we find that the  possibility of observation of $\rho$ meson condensation in weak magnetic field  depends upon the 
  interaction terms used in the lagrangian and undoubtedly demands further investigation.
  
   At this point it is necessary to mention that our phenomenological lagrangian  
 considers only the $\rho\pi\pi$
 interaction which takes into account, in fact, the largest decay channel($\sim$ 100 percent) of rho meson which is $\rho\rightarrow\pi\pi$. Nevertheless, 
 because of its simplicity, 
 our phenomenological  lagrangian can be implemented at 
 finite temperature calculations as well, which will be  discussed elsewhere \cite{tobe}.

 \newpage

\section*{APPENDIX}

Here we list  all the identities necessary to perform  
the momentum integrations of Eq.(\ref{eqnappendics}). All through the section we use $\Delta=x(x-1)k^2+m^2_{\pi}$.

Identity 1:
\bea
\int\frac{d^4p}{(2\pi)^4}\frac{1}{(p^2-\Delta)^5} &=& -i\frac{1}{(4\pi)^2}\frac{1}{(\Delta)^3}\frac{\Gamma[3]}{\Gamma[5]}\nn\\
&=&-\frac{i}{12}\frac{1}{(4\pi)^2}\frac{1}{\Delta^3}
\eea

Identity 2:
\bea
\int\frac{d^4p}{(2\pi)^4}\frac{p^\mu p^\nu}{(p^2-\Delta)^5} &=& \frac{g^{\mu\nu}}{4}\int\frac{d^4p}{(2\pi)^4}\frac{p^2}{(p^2-\Delta)^5}\nn\\
&=&\frac{g^{\mu\nu}}{4}\frac{i}{(4\pi)^2}\frac{1}{(\Delta)^2}\frac{\Gamma[3]}{\Gamma[5]}\nn\\
&=&\frac{i}{48}\frac{1}{(4\pi)^2}\frac{1}{(\Delta)^2}g^{\mu\nu}
\eea
Identity 3:
\bea
\int\frac{d^4p}{(2\pi)^4}\frac{p^2_{||}}{(p^2-\Delta)^5} &=& i\int \frac{d^2p_{\perp}}{(2\pi)^2}
\int\frac{d^2p_{E_{||}}}{(2\pi)^2}\frac{p^2_{E_{||}}}{(p^2_{E_{||}}+\Delta_{||})^5}\nn\\
&=&i\frac{1}{4\pi}\int \frac{d^2p_{\perp}}{(2\pi)^2}\frac{1}{(p^2_{\perp}+\Delta)^3}\frac{\Gamma[3]}{\Gamma[5]}\nn\\
&=&i\frac{1}{(4\pi)^2}\frac{1}{\Delta^2}\frac{\Gamma[2]}{\Gamma[3]}\frac{\Gamma[3]}{\Gamma[5]}\nn\\
&=&\frac{i}{24}\frac{1}{(4\pi)^2}\frac{1}{\Delta^2} 
\eea
Identity 4:
\bea
\int\frac{d^4p}{(2\pi)^4}\frac{p^2_{\perp}}{(p^2-\Delta)^5} &=& -i\int \frac{d^2p_{\perp}}{(2\pi)^2}p^2_{\perp}
\int\frac{d^2p_{E_{||}}}{(2\pi)^2}\frac{1}{(p^2_{E_{||}}+\Delta_{||})^5}\nn\\
&=& -\frac{i}{4\pi}\int \frac{d^2p_{\perp}}{(2\pi)^2}p^2_{\perp}\frac{1}{(p^2_{\perp}+\Delta)^4}\frac{\Gamma[4]}{\Gamma[5]}\nn\\
&=& -\frac{i}{(4\pi)^2}\frac{1}{\Delta^2}\frac{1}{\Gamma[4]}\frac{\Gamma[4]}{\Gamma[5]}\nn\\
&=& -\frac{i}{24}\frac{1}{(4\pi)^2}\frac{1}{\Delta^2}
\eea
Identity 5:
\bea
\int\frac{d^4p}{(2\pi)^4}\frac{p^\mu_{||}p^\nu_{||}}{(p^2-\Delta)^5} 
&=& \frac{g^{\mu\nu}_{||}}{2}\int \frac{d^4p}{(2\pi)^4}\frac{p^2_{||}}{(p^2-\Delta)^5}\nn\\
&=&g^{\mu\nu}_{||}\frac{i}{48}\frac{1}{(4\pi)^2}\frac{1}{\Delta^2} 
\eea
Identity 6:
\bea
\int\frac{d^4p}{(2\pi)^4}\frac{p^\mu_{\perp}p^\nu_{\perp}}{(p^2-\Delta)^5} 
&=& \frac{g^{\mu\nu}_{\perp}}{2}\int \frac{d^4p}{(2\pi)^4}\frac{p^2_{\perp}}{(p^2-\Delta)^5}\nn\\
&=&-g^{\mu\nu}_{\perp}\frac{i}{48}\frac{1}{(4\pi)^2}\frac{1}{\Delta^2} 
\eea
Identity 7:
\bea
\int\frac{d^4p}{(2\pi)^4}\frac{p^2_{||}p^\mu_{||}p^\nu_{||}}{(p^2-\Delta)^5} 
&=& \frac{g^{\mu\nu}_{||}}{2} \int \frac{d^4p}{(2\pi)^4}\frac{p^4_{||}}{(p^2-\Delta)^5}\nn\\
&=& -i\frac{g^{\mu\nu}_{||}}{2}\int \frac{d^2p_\perp}{(2\pi)^2}\int \frac{d^2p_{E_{||}}}{(2\pi)^2}
\frac{(p^2_{E_{||}})^2}{(p^2_{E_{||}}+\Delta_{||})^5}\nn\\
&=& -i\frac{g^{\mu\nu}_{||}}{2}\int \frac{d^2p_\perp}{(2\pi)^2} \frac{1}{4\pi}\frac{1}{(p^2_\perp+\Delta)^2}
\frac{\Gamma[3]}{\Gamma[5]}\nn\\
&=&-g^{\mu\nu}_{||}\frac{i}{24}\frac{1}{(4\pi)^2}\frac{1}{\Delta}
\eea
Identity 8:
\bea
\int\frac{d^4p}{(2\pi)^4}\frac{p^2_{\perp}p^\mu_{||}p^\nu_{||}}{(p^2-\Delta)^5} 
&=& \frac{g^{\mu\nu}_{||}}{2} \int \frac{d^4p}{(2\pi)^4}\frac{p^2_{||}p^2_{\perp}}{(p^2-\Delta)^5}\nn\\
&=& i\frac{g^{\mu\nu}_{||}}{2}\int \frac{d^2p_\perp}{(2\pi)^2}p^2_\perp\int \frac{d^2p_{E_{||}}}{(2\pi)^2}
\frac{p^2_{E_{||}}}{(p^2_{E_{||}}+\Delta_{||})^5}\nn\\
&=& i\frac{g^{\mu\nu}_{||}}{2}\int \frac{d^2p_\perp}{(2\pi)^2}p^2_\perp \frac{1}{4\pi}\frac{1}{(p^2_\perp+\Delta)^3}
\frac{\Gamma[3]}{\Gamma[5]}\nn\\
&=& i\frac{g^{\mu\nu}_{||}}{2}\frac{1}{(4\pi)^2}\frac{1}{\Delta}\frac{1}{\Gamma[3]}\frac{\Gamma[3]}{\Gamma[5]}\nn\\
&=&g^{\mu\nu}_{||}\frac{i}{48}\frac{1}{(4\pi)^2}\frac{1}{\Delta}
\eea
Identity 9:
\bea
\int\frac{d^4p}{(2\pi)^4}\frac{p^2_{||}p^\mu_{\perp}p^\nu_{\perp}}{(p^2-\Delta)^5}
&=&\frac{g^{\mu\nu}_{\perp}}{2}\int\frac{d^4p}{(2\pi)^4}\frac{p^2_{||}p^2_{\perp}}{(p^2-\Delta)^5}\nn\\
&=&g^{\mu\nu}_{\perp}\frac{i}{48}\frac{1}{(4\pi)^2}\frac{1}{\Delta}
\eea
Identity 10:
\bea
\int\frac{d^4p}{(2\pi)^4}\frac{p^2_{\perp}p^\mu_{\perp}p^\nu_{\perp}}{(p^2-\Delta)^5}
&=&\frac{g^{\mu\nu}_{\perp}}{2}\int\frac{d^4p}{(2\pi)^4}\frac{p^4_{\perp}}{(p^2-\Delta)^5}\nn\\
&=&-i\frac{g^{\mu\nu}_{\perp}}{2}\int \frac{d^2p_\perp}{(2\pi)^2}p^4_\perp\int \frac{d^2p_{E_{||}}}{(2\pi)^2}
\frac{1}{(p^2_{E_{||}}+\Delta_{||})^5}\nn\\
&=&-i\frac{g^{\mu\nu}_{\perp}}{2}\int \frac{d^2p_\perp}{(2\pi)^2}p^4_\perp \frac{1}{4\pi}\frac{1}{(p^2_\perp+\Delta)^4}
\frac{\Gamma[4]}{\Gamma[5]}\nn\\
&=&-i\frac{g^{\mu\nu}_{\perp}}{2}\frac{1}{(4\pi)^2}\frac{1}{\Delta}\frac{\Gamma[3]}{\Gamma[4]}\frac{\Gamma[4]}{\Gamma[5]}\nn\\
&=&-g^{\mu\nu}_{\perp}\frac{i}{24}\frac{1}{(4\pi)^2}\frac{1}{\Delta}
\eea
Identity 11:
\bea
&&\int\frac{d^4p}{(2\pi)^4}\frac{p^\mu_{||}(p_{||}\cdot k_{||})}{(p^2-\Delta)^5}\nn\\
&=&\frac{i}{48}\frac{1}{(4\pi)^2}\frac{1}{\Delta^2} k^\mu_{||}
\eea
Identity 12:
\bea
&&\int\frac{d^4p}{(2\pi)^4}\frac{p^\mu_{\perp}(p_{\perp}\cdot k_{\perp})}{(p^2-\Delta)^5}\nn\\
&=&-\frac{i}{48}\frac{1}{(4\pi)^2}\frac{1}{\Delta^2} k^\mu_{\perp}
\eea
Identity 13:
\bea
&&\int\frac{d^4p}{(2\pi)^4}\frac{p^2_{||}}{(p^2-m^2_\pi)^4}\nn\\
&=& -\frac{i}{(4\pi)^2}\frac{1}{6}\frac{1}{m^2_\pi}
\eea
Identity 14:
\bea
&&\int\frac{d^4p}{(2\pi)^4}\frac{p^2_\perp}{(p^2-m^2_\pi)^4}\nn\\
&=& \frac{i}{(4\pi)^2}\frac{1}{6}\frac{1}{m^2_\pi}
\eea
Identity 15:
\bea
&&\int\frac{d^4p}{(2\pi)^4}\frac{1}{(p^2-m^2_\pi)^4}\nn\\
&=& \frac{i}{(4\pi)^2}\frac{1}{6}\frac{1}{(m^2_\pi)^2}
\eea

 \end{document}